\documentstyle[preprint,epsf,12pt,aps,floats]{revtex}
\newcommand{\postscript}[2]{\setlength{\epsfxsize}{#2\hsize} 
   \centerline{\epsfbox{#1}}} 
 
\tighten 
\begin{document} 
\def\te{\tilde e}  
\def\tl{\tilde l}  
\def\tu{\tilde u}  
\def\ts{\tilde s}  
\def\tb{\tilde b}  
\def\tf{\tilde f}  
\def\tt{\tilde t}
\def\td{\tilde d}  
\def\tQ{\tilde Q}  
\def\tL{\tilde L}  
\def\tH{\tilde H}  
\def\tst{\tilde t}  
\def\ttau{\tilde \tau}  
\def\tmu{\tilde \mu}  
\def\tg{\tilde g}  
\def\tnu{\tilde\nu}  
\def\tell{\tilde\ell}  
\def\tq{\tilde q}  
\def\tw{\widetilde W}  
\def\tz{\widetilde Z}  
\draft 
\preprint{\vbox{\baselineskip=14pt%
 \rightline{UH-511-1010-02} 
   \rightline{FSU-HEP-021126}  
}} 
\title{Are Supersymmetric Models with Large $\tan\beta$ Natural?} 
\author{Howard Baer$^1$, Javier Ferrandis$^{2}$ and Xerxes Tata$^{2}$} 
\address{ 
$^1$Department of Physics, 
Florida State University, 
Tallahassee, FL 32306 USA} 
\address{ 
$^1$Department of Physics and Astronomy, 
University of Hawaii, 
Honolulu, HI 96822, USA} 
\date{\today} 
\maketitle 
\begin{abstract} 
We point out that, contrary to general belief, generic supersymmetric
models are not technically unnatural in the limit of very large values
of the parameter $\tan\beta$ when radiative corrections are properly
included. Rather, an upper limit on $\tan\beta$ only arises from the
requirement that Yukawa couplings remain perturbative up to some high
scale. We quantify the relation between this scale and the maximum value
of $\tan\beta$. Whereas $\tan\beta$ is limited to lie below 50--70 in
the mSUGRA model, models with a much lower scale of new physics (beyond
supersymmetry) may have $\tan\beta \alt 150-200$

\end{abstract} 
\medskip 
\newpage 
%
%
%
\section{Introduction} 
%
Although the Standard Model (SM) is spectacularly successful in 
accommodating experimental data over a wide range of energies, it is 
widely believed to be an effective theory that is applicable below an 
energy scale ${\cal O}(1-10)$~TeV. New degrees of freedom (or at least 
evidence of structure via form factors) are expected to manifest 
themselves in experiments designed to probe energies above this scale. 
This belief stems from the instability of the parameters of the 
elementary scalar field sector to radiative corrections. This 
instability, in turn, may be interpreted as an extreme sensitivity of 
weak scale physics to parameters that describe physics at much higher 
energy scales. This is referred to as the fine-tuning problem of the 
SM. We stress that this is not a logical problem in that the SM provides 
an internally consistent predictive framework, but more a problem of what 
we expect of a fundamental theory (which the SM is not). 
 
A conceptually distinct issue refers to the introduction of 
small dimensionless parameters, be they dimensionless couplings or small 
ratios of mass scales, into a theory. It has been proposed\cite{'tHooft} that 
a dimensionless parameter P may be much smaller than unity 
only if the replacement $P\rightarrow 0$ increases the 
symmetry of the theory. Theories that satisfy this requirement are 
technically referred to as natural.  A small Yukawa coupling in 
grand unified theories (GUTs) is technically natural (because setting it to 
zero leads to a new chiral symmetry), 
but the introduction of the tiny ratio of the 
electroweak Higgs boson mass parameter 
to the grand unification scale is not. Likewise, 
within the framework of the simplest supersymmetric GUTs 
the choice $|\mu| \ll M_{GUT}$ is technically unnatural. This is the 
well-known ``$\mu$ problem''. Various dynamical mechanisms have been 
suggested to explain why $\mu$ is of the same size as the SUSY breaking 
scale \cite{mu}.  
 
While supersymmetry, by itself, does not address the naturalness 
question, it has received a lot of attention in the last two decades 
because it leads to an elegant solution to the fine-tuning 
problem\footnote{What we call the fine-tuning problem has been referred 
to as the naturalness problem by some authors.} of the first paragraph, 
provided that the SUSY breaking scale is comparable to the weak 
scale\cite{fine}. Supersymmetry thus preserves the hierarchy between the 
weak and GUT (or Planck) scales even in the presence of radiative 
corrections. But why this hierarchy exists at all requires an 
independent dynamical explanation. 
 
The interpretation of the atmospheric neutrino data of the 
super-Kamiokande collaboration\cite{superk} as neutrino oscillations has 
led to a renewed interest in $SO(10)$ GUTs since neutrinos necessarily 
acquire masses within this framework. SUSY models based on $SO(10)$ 
require that the parameter $\tan\beta$ is large\cite{so10}. It has been argued, 
however, that models with large $\tan\beta$ are technically 
unnatural\cite{nelson}. It is an evaluation of this claim that forms the 
subject of this note. 
 
We begin by examining the part of the scalar potential relevant for  
spontaneous electroweak symmetry breaking (EWSB). At tree-level, this 
takes the form, 
\begin{equation} 
{\cal V}_{tree} = 
\left( m_{H_u}^2+ \mu ^2 \right)  
\left| h_u^{0} \right|^2 +  
\left( m_{H_d}^2+ \mu ^2 \right)  
\left| h_d^{0} \right|^2  
+ g_Z^2  
\left( \left| h_u^{0} \right|^2 - \left| h_d^{0} \right|^2 
\right)^2 
-B \mu \left( h_u^{0}h_d^{0} +h.c. \right),  
\label{eq:potesc} 
\end{equation} 
with 
\begin{equation} 
g_Z^2 = \frac{1}{8}\left( g^2 +g^{\prime 2} \right) =  
\frac{g^2}{8 \cos^2 \theta_W}. 
\label{eq:vev} 
\end{equation} 
%
The minimization conditions can readily be derived from this potential 
and take the well known form, 
%
\begin{eqnarray} 
\mu B &=&  \sin\beta \cos\beta  
\left( m_{H_u}^2 +m_{H_d}^2 + 2 \mu^2  
\right), 
\label{minimizationB1} 
\\ 
\mu^2 
&=& \frac{ m_{H_d}^2 - m_{H_u}^2 \tan^2{\beta}} {(\tan^2\beta -1)} 
- \frac{1}{2}m^2_Z. 
\label{minimizationB2} 
\end{eqnarray} 
%
It follows from (\ref{minimizationB1}) that if $\tan\beta \gg 1$, the 
parameter $B\mu$ has a much smaller magnitude than the other parameters 
in the scalar potential. Thus the model is technically unnatural unless  
the limit $B\mu \rightarrow 0$ (equivalently, $\tan\beta =\infty$) increases 
the symmetry of the Lagrangian.  
Indeed $B$ can naturally be made small by an approximate R symmetry, 
while $\mu$ could be small because of an approximate Peccei--Quinn symmetry, 
which is taken to commute with supersymmetry 
\cite{Hall:1994gn}. 
However, since either of these symmetries requires a chargino with a 
mass below its experimental lower limit,  
an enlargement of the Higgs sector was proposed 
to make the large $\tan\beta$ scenario  
natural\cite{nelson}. 
 
This simple argument is based on an analysis of the vacuum of the 
tree-level potential. In the next section, we show that (unlike at 
tree level) if we take into account radiative corrections to the 
potential, the value of $B\mu$ does not vanish even if $\tan\beta \to 
\infty$. If this radiatively corrected value of $B\mu$ (though loop 
suppressed) is not much smaller than other soft SUSY breaking parameters 
(this could be because of hierarchies of $\sim 10$ in their values, 
which is certainly allowed in a generic SUSY model), the $\tan\beta 
\to \infty$ limit is not unnatural in the sense of Ref.\cite{'tHooft}, 
as implied by the tree-level analysis. This is our main 
point.\footnote{It is common belief that SUSY models are unnatural  
if $\tan\beta$ is large \cite{many}. We believe that this conclusion 
based on the tree level analysis of the vacuum state, for which it 
follows from Eq.~(\ref{minimizationB1}) that $B\mu \to 0$ as $\tan\beta 
\to \infty$. Our point is that the same conclusion does not follow once 
radiative corrections are included.} 
In  
Section~III, we exhibit technically natural scenarios with very large 
values of $\tan\beta$. We note that the upper bound on $\tan\beta$ comes 
from the requirement that Yukawa couplings remain perturbative up to a 
scale $Q_{NP}$, and quantify the relation between the maximum value of 
${\tan\beta }$ 
and this scale. We conclude in the last section with a discussion of our 
analysis.

\section{One loop minimization of the scalar potential} 
%
Radiative corrections cause the ground state of a quantum theory  
to differ from the ground state of the corresponding classical theory. 
These radiative corrections are automatically included when the vacuum 
state is computed by minimizing  
the effective potential  
\begin{displaymath} 
{\cal V} = {\cal V}_{tree} + {\cal V}_{rad}, 
\end{displaymath} 
of the quantum theory, where the  
the one loop radiative correction to the potential, 
renormalized at the scale $Q$, is given by 
\begin{equation} 
{\cal V}_{rad} = \sum_k \frac{1}{64 \pi^2} (-1)^{2J_k} 
\left( 2J_k +1 \right) c_k m_k^4 \left( \log \left( \frac{m_k^2}{Q^2} \right) 
- \frac{3}{2} \right). 
\label{eq:potesceff} 
\end{equation} 
%
Here the sum is taken over independent real boson or Majorana fermion 
fields in the loop (complex boson fields and Dirac fermions, therefore, 
contribute twice as much), $m_k^2$ are the field dependent squared 
masses of the particles in the loops, $J_k$ is their spin, and the 
factor $c_k= 3(1)$ for coloured (uncoloured) particles. 
 
For the minimal supersymmetric Standard Model, the relevant part of  
${\cal V}_{tree}$ is given by (\ref{eq:potesc}). Gauge invariance 
dictates that both ${\cal V}_{tree}$ and ${\cal V}_{rad}$ are functions 
of the field combinations, 
\begin{displaymath} 
|h_u|^2, |h_d|^2 \ {\rm and} \ (h_uh_d+h.c.), 
\end{displaymath} 
so that 
\begin{eqnarray} 
\left. \frac{\partial{\cal V}_{rad}}{\partial h_u^{0*}}\right|_{min}&=& 
\left. \frac{\partial{\cal V}_{rad}}{\partial |h_u|^2}\right|_{min}v_u  
+ \left. \frac{\partial{\cal V}_{rad}}{\partial (h_u h_d + h.c. )} 
\right|_{min}v_d, 
\nonumber \\ 
\left. \frac{\partial{\cal V}_{rad}}{\partial h_d^{0*}}\right|_{min} &=& 
\left. \frac{\partial{\cal V}_{rad}}{\partial |h_d|^2}\right|_{min}v_d 
+ \left. \frac{\partial{\cal V}_{rad}}{\partial (h_u h_d + h.c.)}\right|_{min}v_u. 
\end{eqnarray} 
It is then easy to see that the effect of including the one loop 
correction to the potential is equivalent to the replacements, 
\begin{eqnarray*} 
m_{H_u}^2 &\rightarrow & m_{H_u}^2 + \Sigma_{uu}, \\ 
m_{H_d}^2 &\rightarrow & m_{H_d}^2 + \Sigma_{dd}, \\ 
B\mu & \rightarrow & B\mu -\Sigma_{ud}, 
\end{eqnarray*} 
in the tree level minimization conditions (\ref{minimizationB1}) and 
(\ref{minimizationB2}), 
where, 
\begin{equation} 
\Sigma_{uu} =  
\left. \frac{\partial {\cal V}_{rad}}{\partial |h_{u}|^2} \right| _{min}, 
\quad  
 \Sigma_{dd} =  
\left. \frac{\partial {\cal V}_{rad}}{\partial |h_{d}|^2} \right| _{min}, 
\quad  
\Sigma_{ud} =  
\left. \frac{\partial {\cal V}_{rad}}{\partial (h_{u}h_{d}+h.c)}  
\right| _{min}, 
\label{eq:sigmauuddud} 
\end{equation} 
%
The radiatively corrected minimization conditions are thus given by, 
%
\begin{eqnarray} 
\mu B &=&  \sin\beta \cos\beta  
\left( m_{H_u}^2 + m_{H_d}^2 + 2 \mu^2  
\right) +  \sin\beta \cos\beta  
\left( \Sigma_{uu} + \Sigma_{dd} \right) + \Sigma_{ud}, 
\label{radminimizationB1} 
\\ 
\mu^2 
&=& \frac{ (m_{H_d}^2+\Sigma_d^d) - \tan^2\beta ( m_{H_u}^2+\Sigma_u^u)} 
{(\tan^2\beta -1)} 
- \frac{1}{2}m^2_Z. 
\label{radminimizationB2} 
\end{eqnarray} 
%
In the $\tan \beta \rightarrow \infty $ limit we see that 
%
\begin{equation} 
\mu B =  \Sigma_{ud}, 
\label{largetanmuB} 
\end{equation} 
which, though suppressed by a loop factor, {\it does not vanish}. Indeed as 
long as $\Sigma_{ud}$ is sizeable, models with large $\tan\beta$ do not 
suffer from the naturalness problem.  This is our 
main observation. 
 
The dominant contribution to ${\cal V}_{rad}$ arises from the third 
generation Yukawa interactions. To illustrate solutions with very large 
values of $\tan\beta$, we will ignore electroweak gauge coupling corrections to 
the effective potential which are known to be small, and whose inclusion 
will not significantly affect our results. In this case, $\Sigma_{ud}$ 
that enters the determination of $B\mu$ is given by, 
%
\begin{displaymath} 
 \Sigma_{ud} =   \Sigma_{ud} (\tilde{t}) + \Sigma_{ud} (\tilde{b}) + 
 \Sigma_{ud} (\tilde{\tau}), 
\end{displaymath} 
with  
%
\begin{equation} 
 \Sigma_{ud} (\tilde{f})=  
\frac{c_{f}}{16 \pi^2}(-\mu) f^2_f A_{f}  
\frac{\left( f(m_{\tilde{f}_2}^2)-f(m_{\tilde{f}_1}^2) 
\right)}{\left(m_{\tilde{f}_2}^2 - m_{\tilde{f}_1}^2 \right)}\;. 
\label{eq:sigudsqlp} 
\end{equation} 
%
Here, $f=t,b,\tau$, $f_f$ is the Yukawa coupling of fermion $f$, 
${\tilde{f}}_i$ are the sfermion mass eigenstates, 
the colour factor $c_f = 3(1)$ when $f$ is a quark (lepton), and  
the function $f(x)$ that appears in (\ref{eq:sigudsqlp}) is given by, 
\begin{displaymath} 
f(x) = x \left(\ln\frac{x}{Q^2} - 1\right)\;. 
\end{displaymath} 
In a generic SUSY model, it is entirely possible that the weak scale 
$A$-parameters and $|\mu|$ are a few times larger than the soft SUSY 
breaking masses; in this case, the factor ${{\mu A_f}\over m_{\tilde{f}}^2}$ 
would largely compensate for the loop suppression, and $\Sigma_{ud}$ 
(and hence, $B\mu$) would be comparable to other soft SUSY breaking 
masses.  In any case, fine-tuning of parameters at the level of ${\cal 
O}({1\over{\tan\beta}})$, that is suggested by the tree-level analysis, 
is not needed. 
  
\section{Very large $\tan\beta$ and the scale of new physics} 
%
 
Although we have argued that very large 
$\tan\beta$ solutions are not necessarily unnatural within 
the MSSM framework, it still remains to be shown that 
those solutions can be phenomenologically viable 
and theoretically interesting. By this, we mean that we look for 
large $\tan\beta$ solutions with third generation 
matter fermion masses given by their  
experimental values, and with the corresponding Yukawa couplings 
in the perturbative range.\footnote{We ignore Yukawa couplings and 
fermion masses for the first two generations.}

In a supersymmetric theory,  
the experimental values for the third 
generation fermion masses determine the corresponding  
Yukawa couplings at $Q=M_Z$, 
but only if we know the sparticle mass 
spectrum. This is because supersymmetric particle loops affect  
the fermion masses through threshold corrections \cite{bmass}. 
Analytical expressions for the one loop susy  
threshold corrections were given in the literature 
\cite{Pierce:1996zz}. To leave our approach as model-independent as  
possible, we parametrize our ignorance of 
the supersymmetric threshold corrections to third generation fermion masses 
through a set of coefficients, $\delta_t, \delta_b, \delta_{\tau}$,  
which appear in the relations between matter fermion masses and the  
corresponding Yukawa couplings: 
%
\begin{eqnarray} 
f_t &=& \frac{m_t}{v \sin\beta}\frac{1}{\left(1+\delta_{QCD} +  
\delta_t \right)}, \nonumber \\  
f_b &=& \frac{m_b(m_Z)}{v \cos\beta}\frac{1}{\left(1+\delta_b  
\tan\beta\right)}, \label{eq:thirdyuks} \\ 
f_{\tau} &=& \frac{m_{\tau}(m_Z)}{v 
\cos\beta}\frac{1}{\left(1+\delta_{\tau}  
\tan\beta \right)}, \nonumber 
\end{eqnarray} 
%
where $v=\sqrt{v_u^2+v_d^2}$. Here, $m_t$ is the top quark pole mass while 
$m_b(Q)$ and $m_{\tau}(Q)$ are running bottom quark and tau lepton 
masses at the scale $Q$ in the $\overline{DR}$ scheme. The coefficient 
$\delta_{QCD}$ is the usual QCD correction relating the pole and running top 
masses. This does not appear in the formula for $f_b$ since for this 
we use the running mass $m_b(M_Z)=2.83$~GeV as the experimental input 
\cite{melnikov}. 
The appearance of $\tan\beta$ in the expressions for bottom and tau 
Yukawa couplings captures the fact that, for large $\tan\beta$,  
the SUSY corrections to $m_b$ 
and $m_{\tau}$ scale with $\tan\beta$ \cite{bmass,Pierce:1996zz}.  
As we have already noted, the SUSY threshold corrections depend on the 
unknown sparticle spectrum. In the following, we implement these by adopting   
reasonable values of the coefficients $\delta_t$, $\delta_b$ and 
$\delta_{\tau}$ as given in typical models with sparticles in the range  
of 100-1000~GeV. Specifically, $\delta_t$  is positive, and increases 
logarithmically with $m_{\tilde{g}}$. Our results are insensitive to its 
precise value which we take to be  0.04, corresponding to $m_{\tilde{g}}$ somewhat 
larger than 1 TeV. The threshold corrections for the down type fermions 
 depend on $\mu$, and so can have either sign.  
We take $\delta_b= \pm 0.008$ which gives a susy threshold of $40\%$ for 
$\tan\beta=50$, typical of the mSUGRA framework with TeV scale  
parameters. For $\delta_{\tau}$, we take $\pm 0.0016$ which correspond 
to $\pm 8\%$ for $\tan\beta=50$ as a typical 
value, and 
$\pm 0.003$ as a somewhat extreme case.

\begin{figure}[tb]       
\postscript{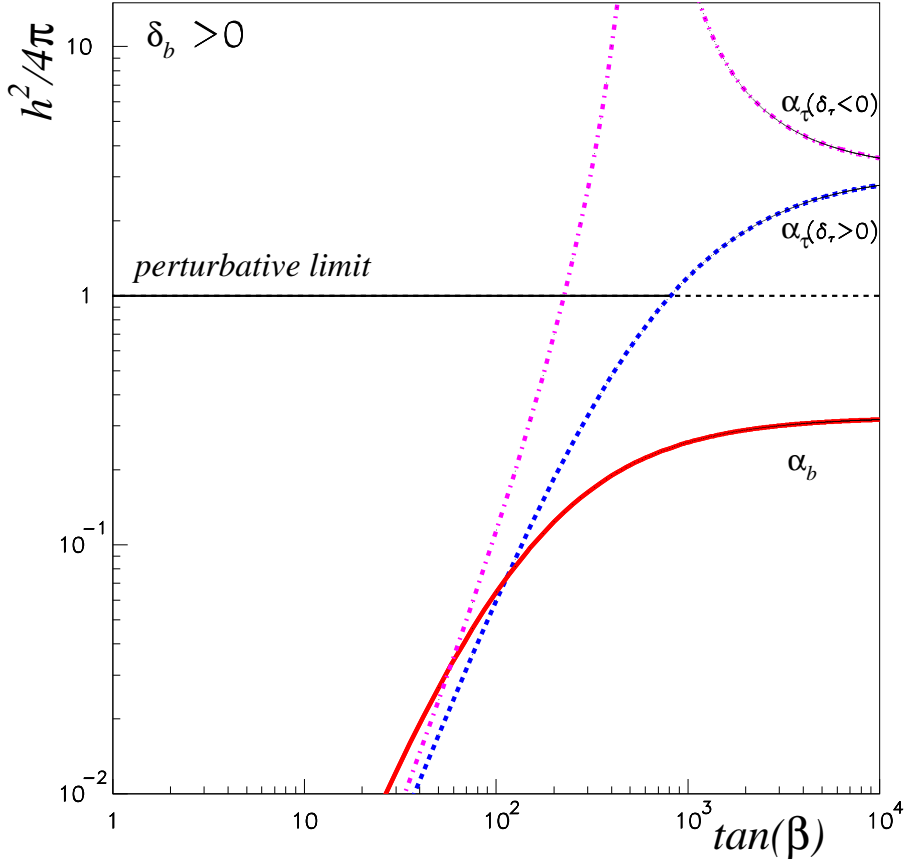}{0.80}       
\caption{\it Bottom and tau Yukawa couplings, $\alpha_b=h^2_b/4\pi$ 
and $\alpha_{\tau}=h^2_{\tau}/4\pi$, 
at $m_Z$ as a function  
of $\tan\beta$ for $\delta_b=+0.008>0$ and two values of $\delta_{\tau}$: 
$+0.0016$ and $-0.0016$} 
\label{fig:yuksp}       
\end{figure}       
 
If all other things are the same, the bottom Yukawa coupling is clearly 
larger if $\delta_b < 0$. Thus the largest values of $\tan\beta$ for 
which $f_b$ remains in the perturbative range occur when $\delta_b 
> 0$ (assuming $f_b$ is positive). Likewise, we would expect that the 
requirement that $f_{\tau}$ lie in the perturbative range would allow 
larger values of $\tan\beta$ when $\delta_{\tau}>0$; however, if 
$\delta_{\tau}$ is negative, for $\tan\beta > {1\over \delta_\tau}$, 
$f_{\tau} < 0$, so that $\frac{f_{\tau}^2}{4\pi}$ decreases as 
$\tan\beta$ increases beyond this value. It is easy to check that this 
branch of the $\delta_{\tau} < 0$ curve asymptotically approaches the 
$\delta_\tau > 0$ curve, as shown in Fig.~\ref{fig:yuksp} where we show 
the value of $\alpha_f \equiv \frac{f_f^2}{4\pi}$ evaluated at $Q=M_Z$ 
versus $\tan\beta$ for $\delta_b=0.008$ and $\delta_{\tau}=\pm 
0.0016$. Naively, one might conclude that ``viable'' solutions to the 
MSSM are possible for $\tan\beta$ values up to 800. 
%
\begin{figure}[tb]       
\postscript{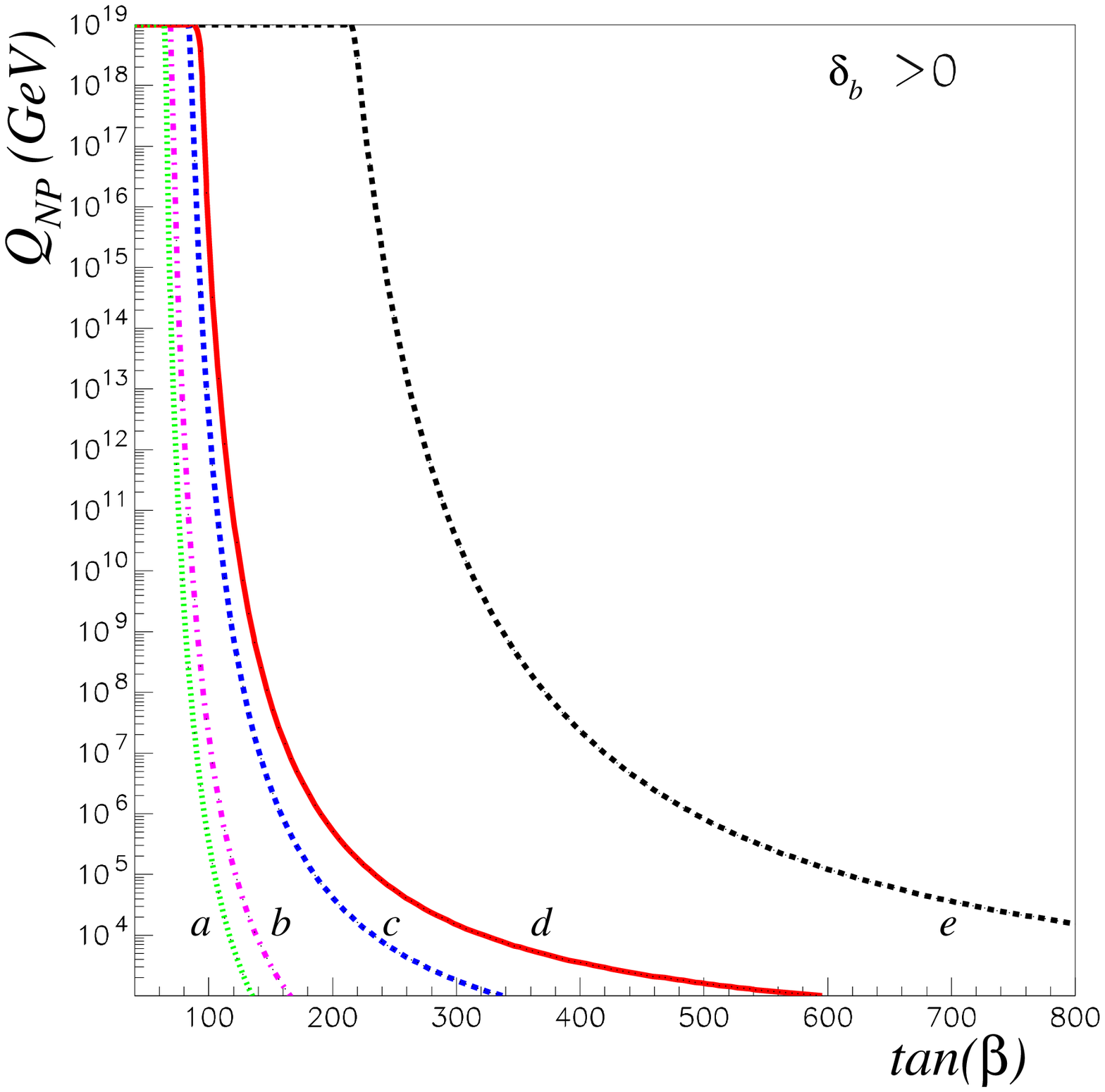}{0.80}       
\caption{\it Relation between $\tan\beta$ and the scale of new physics 
for $\delta_b=+0.008>0$ and four values of  
$\delta_{\tau}$: $-0.003$ (dotted--green line, labelled a), 
 $-0.0016$ (dash/dotted--pink line, labelled b) 
$+0.0016$ (dashed--blue line, labelled c) and 
$+0.003$ (solid--red line, labelled d). The dashed black line labelled e 
shows the value of $Q_{NP}$ for a fictitious case with both $m_{\tau}$ 
and $m_b$ set at one half their experimental values.} 
\label{fig:natural}       
\end{figure}       
%

This is, however, not the case since a Yukawa coupling close to the 
``perturbative limit'' (aside from the fact that this might be 
phenomenologically unacceptable) would blow up at a scale $Q_{NP}$ not 
far above $M_Z$, and we would lose our main motivation for {\it weak 
scale} supersymmetry. Weak scale supersymmetry is well-motivated only if 
the scale $Q_{NP}$ where {\it any} coupling becomes non-perturbative is 
sufficiently separated from $M_Z$. We stress that the electroweak scale 
is destabilized even if $Q_{NP}$ is much smaller than $M_{Planck}$ or 
$M_{GUT}$: it is both model-dependent and a matter of judgement just how 
large a $Q_{NP}$ is acceptable in any extension of the SM without 
supersymmetry to stabilize the electroweak scale. Since a discussion of 
this would take us away from our main point, we show the value of 
$Q_{NP}$ as a function of $\tan\beta$ in Fig.~\ref{fig:natural}. We have 
again fixed $\delta_b =0.008$ and illustrate our result for 
$\delta_{\tau}= -0.003$ (dotted green curve, labelled a), 
$\delta_{\tau}= -0.0016$ (dashed-dotted pink line, labelled b), 
$\delta_{\tau}= 0.0016$ (dashed blue line, labelled c) and 
$\delta_{\tau}= 0.003$ (solid red line, labelled d). In all these cases, 
it is the coupling $\alpha_{\tau}$ that exceeds unity at $Q=Q_{NP}$, 
with the other Yukawa couplings remaining perturbative. 
This figure, which  updates previous work by
Haber and Zwirner \cite{Haber:1993pv},
includes two-loop Yukawa coupling RGEs and models the
effect of SUSY threshold corrections. 
We see that for 
$Q_{NP}$ close to $M_{GUT}$, the maximum value of $\tan\beta$ is never 
much above what one obtains within the mSUGRA framework. We stress 
though that the value of this maximum is dictated by the  
measured fermion masses and {\it not by 
the naturalness considerations}. To emphasize 
this, we show the corresponding curve (dashed black line, labelled e) 
for $\delta_{\tau}=0.0016$ but with tau lepton and bottom quark masses 
fixed at half their experimental values. In such a universe, it is 
easily  
possible to find natural models with $\tan\beta$ larger than 200, and 
couplings in the perturbative range all the way up to 
$Q=M_{Planck}$. Returning to the case of realistic masses, we 
see from the figure that (depending on the value of SUSY thresholds) 
models with $\tan\beta$ as large as 150 may be natural if the new 
physics scale is smaller than $\sim 10^7$~GeV, and we use SUSY to 
stabilize the electroweak scale relative to this intermediate 
scale.\footnote{We recognize that we would still have to be careful 
about the new physics at this scale in order not to have to attribute 
the apparent unification of gauge couplings measured by LEP experiments 
to an accident.} 
 
It is well known that SUSY phenomenology of large $\tan\beta$
models differs considerably \cite{large} from that of models with low or
moderate values of $\tan\beta$. What is less clear (because in the
well-studied models, $\tan\beta \alt 50-70$) is whether the
phenomenology is altered as $\tan\beta$ is changed from $\sim 50$ to $\agt
100$. As we have already explained, $\tan\beta \ge 100$ can only be
accommodated if $Q_{NP}$ is relatively low. This led us to examine the
gauge-mediated SUSY breaking (GMSB) framework with a low messenger
scale. Within the minimal version of this model, the radiative
electroweak symmetry breaking mechanism breaks down, leading to $m_A^2 <
0$ (for small $\Lambda$ values, $m_{\ttau_1}^2 < 0$), if $\tan\beta$ is
too large.\footnote{Within this framework, Higgs boson mass squared
parameters are mainly driven down because squarks are heavy. If
$\tan\beta$ is very large, the bottom and top Yukawa couplings are
similar, so that $m_{H_u}^2$ and $m_{H_d}^2$ start from a common value
and roughly evolve together. This then causes $m_A^2 \simeq
m_{H_d}^2-m_{H_u}^2-M_Z^2$ (valid for large values of $\tan\beta$) to
turn negative.} To obtain larger values of $\tan\beta$, we introduced
additional contributions $\delta m_{H_{u,d}}^2$ to the soft SUSY
breaking Higgs boson masses, since these can facilitate radiative
electroweak symmetry breaking. We attribute their origin to additional
interactions needed to generate the $\mu$ and $B\mu$ parameters within
this framework~\cite{thomas}.
\begin{table}
\begin{center}
\caption{Selected sparticle masses 
for two non--minimal GMSB scenario with $M_{mess}=2\Lambda= 300$~TeV,
$n_5=2$, $\mu>0$ and
$\tan\beta =50$ and $\tan\beta =100$. We take $\delta m_{H_u}^2 
= -(1000 \ {\rm GeV})^2$ and $\delta m_{H_d}^2 =0$. For the
corresponding minimal model, the upper limit on $\tan\beta$ is about
68. The entry in the last column is $10^9 \times B(B_s \to \mu^+\mu^-)$. }
\bigskip
\begin{tabular}{|l|cccccccc|}
$\tan\beta$  & $m_{\ttau_1}$ & $m_A$ & $m_{\tt_1}$ &  $m_{\tb_1}$
& $m_{\te_R}$  & $m_{\tu_L}$ & $m_{\tz_1}$ & $B_s \to \mu^+\mu^-$ \\ \hline
$50$ & $275$ & $1245.4$ & $2046.7$ & $2102$ & $385$ & $2289$ & 
$425.4$ & 4.3 \\
$100$ & $99.9$ & $419.7$ & $2038.4$ & $1908$ & $386$ & $2286$ &  
$425.5$ & 33
\label{nonminGMSB}
\end{tabular}
\end{center}
\end{table}
We have used ISAJET v7.64 \cite{isajet} to evaluate the mass spectrum
for two scenarios, with $\tan\beta=50, 100$ whose parameters are
listed in Table~\ref{nonminGMSB}, where selected sparticle masses
are shown. In both scenarios, $\ttau_1$ is the second lightest sparticle.
The sfermions of the first two generations and the charginos and
neutralinos have the same masses to within about a percent in the two
cases. Third generation squark masses differ by a fraction of a percent
for $t$-squarks to about 10\% for $\tb_1$. The most striking difference
is in the mass of the $A$ (and associated $H$ and $H^{\pm}$), and the
mass of the lighter stau. The value of $m_{\te_L}/m_{\te_R}$ would
suggest a GMSB scenario; the relative lightness of $\ttau_1$ would 
point to the very large value of $\tan\beta$.~\footnote{Remember that
$m_A$ always becomes small near the upper bound of $\tan\beta$, so by
itself would not be indicative of the value $\tan\beta$.}
Tevatron experiments may be able to probe the
$\tan\beta=100$ scenario in the Table via the decay $B_s \to \mu^+\mu^-$ whose
branching fraction is usually very small in GMSB models
\cite{mizukoshi}. We have checked that $B(b\to s\gamma)$ is similar in
both cases: 3.44 (3.95)$\times 10^{-4}$ for $\tan\beta = 50$~(100),
while $\Delta a_{\mu} = 11$ (22)$\times 10^{-10}$ scales with
$\tan\beta$ as expected.
The message of this illustrative
example is that changing $\tan\beta$ from a large to a very large value
has experimentally interesting
implications.

\section{Discussion} 
 
We have pointed out that the usual arguments \cite{nelson} that suggest 
that supersymmetric extensions of the SM are unnatural for large values 
of the parameter $\tan\beta$ are inapplicable when 1-loop corrections to 
the effective potential are included: as can be seen from  
(\ref{largetanmuB}), the soft mass parameter $\mu B$ does not vanish as 
$\tan\beta \rightarrow \infty$, and the question of checking whether 
there is an increased symmetry when $\mu B \rightarrow 0$ becomes moot. 
In a  
{\it generic} SUSY model, it appears to us that there is no naturalness 
problem in the sense discussed in Ref.~\cite{'tHooft}.  
 
We stress though that there may be a different fine-tuning required in specific 
models if $\tan\beta$ is very large. For instance, in the mSUGRA 
framework, characterized by the soft breaking parameters, $m_0, m_{1/2}, 
A_0, (B\mu)_0$ and $\mu_0$ at the high scale, the parameter $B\mu$ at 
the weak scale will be of loop-suppressed magnitude only if the high 
scale parameters are all of comparable size, but related in a specific 
manner. From the perspective of a low energy theorist who does not have 
an understanding of the SUSY breaking mechanism, this appears to require 
an unexplained adjustment of the underlying parameters.\footnote{Of 
course, the necessary ``fine-tuning'' is much less severe than for the 
SM Higgs mass parameter because the underlying supersymmetry still 
precludes quadratic divergences in the corrections to $B\mu$.} 
But, perhaps, it is better to 
view this as a necessary property of the physics underlying 
supersymmetry breaking; {\it i.e.} the soft parameters that emerge from 
the theory of supersymmetry breaking must be related so that $B\mu$ is 
small at the weak scale. However, this issue seems to be separate from 
the naturalness question that we have focussed upon. 
 
In summary, it appears to us that SUSY models are not unnatural in the
technical sense of 't Hooft even if the parameter $\tan\beta$ is large.
We have argued that the parameter $B\mu$ does not vanish in the limit
$\tan\beta \rightarrow \infty$ when radiative corrections are included.
Our considerations could be especially relevant in the context of low
scale supersymmetric models as GMSB models \cite{Giudice:1998bp} or
supersymmetric extra dimensional models which are the subject of recent
interest \cite{susywarp}, and for which new physics (beyond
minimal SUSY) intervenes at a relatively low scale. 
%
\acknowledgements 
%
We thank M. Drees for discussion at an early stage of this 
work.  This research was supported in part by the U.S. Department of 
Energy under grants DE-FG02-97ER41022 and DE-FG03-94ER40833. 
%
 
%
\end{document}